\documentclass[prl,amsmath,amssymb,floatfix,twocolumn]{revtex4}%
\usepackage{graphicx,color,epsfig,epstopdf}
\usepackage{bm}

\newcommand{\beq}{\begin{eqnarray}}
\newcommand{\eeq}{\end{eqnarray}}

\newcommand{\D}{{\rm d}}

\def \be{\begin{equation}}
\def \ee{\end{equation}}
\def \ba{\begin{array}}
\def \ea{\end{array}}
\def \bea{\begin{eqnarray}}
\def \eea{\end{eqnarray}}

\def \l{\left}
\def \rr{\right}

\begin{document}

\title{Supercurrent survival under Rosen-Zener quench of hard core bosons}

\author{I. Klich$^{1}$, C. Lannert$^{2}$, and G. Refael$^{1}$}%
\affiliation{(1) Department of Physics,
California Institute of  Technology, MC 114-36 Pasadena, CA 91125
\\ (2) Department of Physics, Wellesley College, Wellesley, MA 02481, USA }

\begin{abstract}
We study the survival of super-currents in a system of
impenetrable bosons subject to a quantum quench from its critical superfluid phase to
an insulating phase. We show that the
evolution of the current when the quench follows a Rosen-Zener profile
is exactly solvable. This allows us to analyze a quench of arbitrary
rate, from a sudden destruction of the superfluid to a slow opening
of a gap. The decay and oscillations of the current are
analytically derived, and studied numerically along with the momentum
distribution after the quench. In the case of small supercurrent boosts $\nu$, we
find that the current surviving at long times is proportional to $\nu^3$.
\end{abstract}

\maketitle


Progress in the field of experimental cold atom systems allows a
controlled and direct access to the non-equilibrium physics of
quantum many body systems. This becomes particularly exciting when
the system is in the vicinity of a phase transition and is driven
through it as in the case of the superfluid(SF)-insulator
transition \cite{Tuchman et al}, or the magnetic ordering
transition \cite{Stamper-Kurn}. In such situations, complex
physics is often exhibited; a prominent example is defect
formation in the ordered phase due to a fast quench \cite{Maniv
Polturak Koren03}, which is qualitatively understood using the
Kibble-Zurek mechanism \cite{Kibble Zurek} originally proposed in
cosmology to describe domain formation during cooling of the
universe. Especially interesting are low-dimensional quantum
systems under a ``quantum quench", with a Hamiltonian driven
through a {\it quantum-critical} point. This can be achieved in
quantum gases confined in highly anisotropic traps and in optical
lattices \cite{experiments,HCB realization}. For instance, it was
recently demonstrated that when a system is driven from an ordered
to a disordered phase, the order-parameter correlations experience
`revival'\cite{Tuchman et al,Polkovnikov et al05,Altman Auerbach
02,Cherng Levitov,Rigol Muramatsu}. Also, in the special case of
driving an off-critical 1d system into criticality, Calabrese and
Cardy \cite{Calabrese Cardy} showed that the long-time
behavior of correlation functions can be obtained from the
correlations in the final critical state. Despite these advances,
as well as qualitative understanding of `revival' phenomena
and the Kibble-Zurek mechanism, analytical and exact results in
this field are scarce.

In this manuscript we focus on a system driven out of criticality
by the variation of an external field (as opposed to Ref.
\cite{Calabrese Cardy}, where a system is brought {\it to}
criticality). In practice, this is a generic case, but elegant
general results as in Ref. \cite{Calabrese Cardy} for the opposing
case are by and large absent. Special cases that have previously
been studied analytically are the behavior of a dipole model of a
Mott insulator in an external electric field when the field is
suddenly changed from a ``no dipole" state to a maximally
polarized state \cite{Sengupta Powell Sachdev}, and the dynamics
of traversing spin chains between two phases \cite{Cherng
Levitov}. A related work, Ref. \cite{Altman Auerbach 02},
describes bosons driven abruptly from a Mott to superfluid phase
and showed collective oscillations of the superfluid order
parameter with period proportional to the gap in the initial Mott
state. In \cite{Rigol Muramatsu}, the evolution of hard-core bosons
(HCBs) undergoing a quantum quench was studied numerically.

A natural question arising in the SF-insulator transition regards the
fate of supercurrents in the system.  In this manuscript we describe in detail the evolution
of supercurrents under quenching a 1-d system of HCBs, the so called Tonks-Girardeu gas
\cite{Tonks
  Girardeau,Lenard}, from superfluid to insulator. Such systems may be
formed by bosons at low temperatures and densities \cite{Olshanii,
HCB realization}. We calculate the current as a function of time,
while concentrating on the long-time current survival rate. Our
analysis reveals fast, Bloch like, oscillations that are
superimposed on a decay and a slower envelope function. The
surviving current is found to be proportional to $\nu^3$, where
$\nu$ is the initial supercurrent. Note that the decay of
supercurrents was considered before mainly in Ref.
\cite{Polkovnikov
  et al05} close to the Mott-superfluid transition. In addition, we present numerics
  for the evolution of the momentum distribution after a
quench in the presence of a supercurrent.

Our study of the supercurrent quench dynamics relies on a
generalization of the Rosen-Zener problem \cite{Rosen Zener} to
the context of HCBs. The Rosen-Zener problem describes a spin
evolving in a time-dependent magnetic field with a particular
profile, where an $x$-$y$ magnetic field (analogous to a gap) is turned
on while a $z$-field remains constant. In contrast, Landau-Zener
like dynamics describe a spin in a constant $x$-$y$ field, with
$B_z$ swept {\it through} zero; thus it is more useful for
describing traversing the system {\it through} a quantum critical
point \cite{Cherng Levitov}. The integrability of the Rosen-Zener
evolution allows us to probe sudden quench dynamics as well as the
response to a finite quenching time and so goes beyond previous
treatments, which have dealt with an abrupt quench.

The system we consider has the Hamiltonian
\begin{eqnarray}
H=-w\sum (b^{\dag}_ib_{i+1}+h.c.)+V(t)\sum (-1)^i b^{\dag}_ib_{i}
\end{eqnarray}
with $b^{\dag}_i$ a boson creation operator at site $i$ which
obeys $(b^{\dag}_i)^2=0$, thus imposing the impenetrability of the
HCBs. Throughout, we work in units such that the lattice spacing is 1.
$V(t)$ is the amplitude of an externally applied potential,
which tunes the system from its superfluid phase at $V=0$, to an
insulating phase at $V\neq 0$. When $V(t)$ has the so-called
Rosen-Zener shape, this evolution can be solved exactly. The
Rosen-Zener profile is given by:
\begin{eqnarray}\label{Rosen Zener delta}
  V(t)=\left\{%
\begin{array}{ll}
     V_0{1\over \cosh {\pi t\over T}} ; & \hbox{$t<0$}, \\
    V_0; & \hbox{$t\geq 0$}, \\
\end{array}%
\right.
\end{eqnarray}
where $T$ is the turning-on, or quenching time.

\begin{figure}
\includegraphics[scale=0.8]{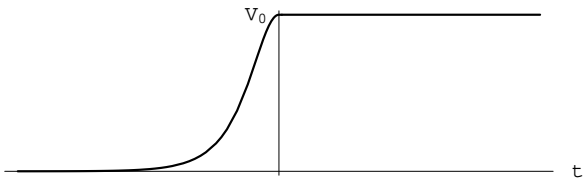}
\caption{The Rosen-Zener switch-on profile $V(t)$ for the
  super-lattice.
\label{Rosen Zener Switch Figure}}
\end{figure}

As we shall see, the HCB problem is equivalent to an infinite
system of spins precessing in a time-dependent magnetic field.
The evolution of a spin in the $\hat{z}$ direction under
$B_x=V(t)$ given in Eq. \eqref{Rosen Zener delta} (fig. \eqref{Rosen
Zener Switch Figure}) was solved exactly by Rosen and Zener
\cite{Rosen Zener}. This exact non-equilibrium evolution allows us to establish the state of the system at $t=0$, from
which the system is evolved by the final Hamiltonian. We make use of
this evolution to analyze the fate of a supercurrent, introduced at
$t\rightarrow-\infty$ (when the system is in the gapless
SF phase and $V(-\infty)=0$) by choosing our initial
state as the ground state of the boosted Hamiltonian $-w\sum
(e^{i\nu}b^{\dag}_ib_{i+1}+h.c.)$.

The first step in our analysis applies the Wigner-Jordan
transformation to the HCB operators \cite{Lieb
Schultz Mattis},
\begin{eqnarray}
b^{\dag}_i=e^{\sum_{j<i} a^{\dag}_ja_j}a^{\dag}_i
\label{cdef}
\end{eqnarray}
with $a_j$ being a fermion operator associated with site $j$. In
the ``fermion" picture, the presence of the super-current is
expressed by choosing the ``Fermi sea" of the $a$ fermions to be
shifted in momentum. That is, the $k$ occupation of the fermions
is
\begin{eqnarray}\label{Fermi step supercurrent}
n_s(k)=\Theta(-k_F+\nu< k < k_F+\nu),
\end{eqnarray}
where $k_F$ is the Fermi
momentum. In the following we will assume half filling, so $k_F=\pi/2$.
The current density is given by:
\begin{eqnarray}\label{current definition}
j={iw\over L}\sum_l  \langle b^{\dag}_{l+1}b_l-h.c. \rangle
\end{eqnarray}
where $L$ is the total number of sites.

To find the evolution of the current we first address the
evolution of the fermion operators introduced in Eq. (\ref{cdef}).
By rewriting the Hamiltonian as:
\begin{eqnarray}&
  H_0= -\sum_{k<|\pi|}
  2w\cos(k)a^{\dag}_ka_k \\ \nonumber & H_d={V(t)\over 2}\sum_{k<|\pi|}(
 a^{\dag}_ka_{k+\pi}+a^{\dag}_{k+\pi}a_k)
\end{eqnarray}
we note that $k$ couples to $k+\pi$ (or equivalently $k-\pi$ since
$k$ is the same up to multiples of $2\pi$). Thus we can write the
Hamiltonian as:
\begin{eqnarray}
  H(t)=\otimes_{|k|<{\pi\over 2}} H_k(t)
\end{eqnarray}
\begin{eqnarray}\label{H k}
\mbox{with:} \hspace{1cm} H_k(t)=2w\cos(k)\sigma_z-{V(t)}\sigma_x
\end{eqnarray}
acting in the $\{a^{\dag}_k,a^{\dag}_{k+\pi}\}$ mode space, and
explicitly breaking the problem into a product of noninteracting spin-systems. Thus
solving the evolution of the HCB system is equivalent to solving for the time
evolution of an infinite series of time-dependent two-level
systems.

For a given $k$, the Schr\"odinger equation generated by Eq.
\eqref{H k} for a fermion operator
$\psi^{\dag}(t)=s(t)a^{\dag}_k+p(t)a^{\dag}_{k+\pi}$ can be
rewritten as a second order differential equation of the form:
\begin{eqnarray}
\ddot{S}=-{V_0^2\over \cosh({\pi t\over T})^2} S+(4iw\cos(k)-{\pi
\over T}\tanh({\pi t\over T}))\dot{S}
\end{eqnarray}
for $S(t)=e^{2iw\cos(k)t}s(t)$, and the same equation for
$P(t)=e^{-2iw\cos(k)t}p(t)$, with $w\rightarrow -w$. This
equation, via a change of variable $z={1\over 2}(\tanh({\pi t\over
T})+1)$, can be recast into a hyper-geometric differential
equation. The solution, satisfying the initial condition
$|S(-\infty)|=1,\,|P(-\infty)=0|$ (recall $|k|<\frac{\pi}{2}$), is
given in terms of the hyper-geometric function: \be\label{S and P}
\ba{c}
  S(z)= {}_2F_1(\alpha,-\alpha,c;z)\\
  P(z)= -i\sqrt{z(1-z)}{2\pi\alpha^2e^{-4iw\cos(k)t}\over c T}\times
  \\ {}_2F_1(1+\alpha,1-\alpha,1+c;z)
\ea
\ee
where \footnote{One may also include dissipation, which amounts to
taking $c\rightarrow c+{\gamma T\over \pi}$, where $\gamma$ is
related to the decay of the higher energy level, which we shall
ignore here \cite{Robiscoe}.}:
\begin{eqnarray}
  \alpha={V_0T\over \pi}\,\,\,\,;\,\,\,\, c={1\over
  2}-{2iwT\cos(k)\over \pi}
\end{eqnarray}

We now use the solution at $t=0$ (by setting $z\rightarrow1/2$ in
\eqref{S and P})  as a boundary condition for the dynamics under
the final Hamiltonian.


The evolution of the fermion operators at times $t>0$ is obtained
by diagonalizing Eq. \eqref{H k} at fixed $V=V_0$. This results in:
\begin{eqnarray}\label{evolution of fermion
operators}&
a^{\dag}_k(t)=A_ka^{\dag}_k(0)+B_ka^{\dag}_{k+\pi}(0)\\ &
a^{\dag}_{k+\pi}(t)=\bar{A_k}a^{\dag}_{k+\pi}(0)-B_ka^{\dag}_k(0)
\end{eqnarray}
where $A,B$ are given explicitly by the relations:
\begin{eqnarray}&
  A_k=\cos \l(E_{k}t\rr)+i\cos \theta\sin \l(E_k\rr) \\ &
  B_k=i\sin\theta\sin \l(E_k t\rr)
\end{eqnarray}
 with $-\pi<k<\pi$,
\begin{eqnarray}\label{E k}
  E_{k}= \sqrt{4w^2\cos^2k+{V}_0^2}
\end{eqnarray}
and $\theta$ is defined through
\begin{eqnarray}
\tan\theta=-{V_0\over 2w\cos k}.
\end{eqnarray}
Using Eq. \eqref{S and P} with $z\rightarrow 1/2$ to find
$a^{\dag}_k(0),a^{\dag}_{k+\pi}(0)$ and substituting in Eq.
\eqref{current definition}, we find the current as a function of
time
\begin{eqnarray}&\label{current integral}
\langle j\rangle=\\
\nonumber & 2wa\int_{-\pi/2}^{\pi/2} {\D k\over 2\pi}\sin(ka)
(2|s_0A_k-p_0B_k^*|^2-1)(n_k-n_{k+\pi})
\end{eqnarray}
where $n_k$ is the fermion momentum (Fermi-Dirac) occupation:
$\langle a^{\dag}_ka_{k'}\rangle=\delta_{kk'}n_k$. The above
equations \eqref{S and P}-\eqref{current integral} constitute a
complete solution to the problem of the current
evolution under a Rosen-Zener quench. 
Below we focus on some interesting special cases.

In particular, we concentrate on an initial state with a super-current at
half filling and with a short switching time, $w T \nu\ll 1$,
 which allows non-adiabatic transitions. In this
case, to leading order near $k\sim\pi/2$, we may take $c\sim 1/2$,
$\Gamma(c)\approx\Gamma(1/2)=\sqrt{\pi}$, giving:
\begin{equation}
\begin{array}{cc}
\label{p0 and s0}
 p(0)=-i\sin\l({V_0 T\over 2 }\rr) & s(0)=\cos\l({V_0T\over2}\rr).
\end{array}
\end{equation}
Assuming small supercurrents ($\nu \ll 1$), the current evolution,
Eq. \eqref{current integral},  simplifies to:
\begin{eqnarray}
\label{current envelope Rosen Zener}
 \langle j \rangle={2aw\nu\over \pi}{\cal M}((2t-T)V_0, \kappa \nu \sqrt{t})\\
\label{Def M}
\hspace{-0.5cm}\mbox{where:}\hspace{0.5cm} {  \cal M}(x,y)={\cos(x){\cal C}(y)-\sin(x){\cal S}(y)\over y}
\end{eqnarray}
and ${\cal S,C}$ are the Fresnel functions  ${\cal
S}(z)=\int_0^z\D u\sin({\pi\over 2} u^2)$,  ${\cal
C}(z)=\int_0^z\D u\cos({\pi\over 2} u^2)$, and $\kappa=\sqrt{{8
w^2\over
  \pi V_0}}$.

\begin{figure}
\includegraphics[width=8cm]{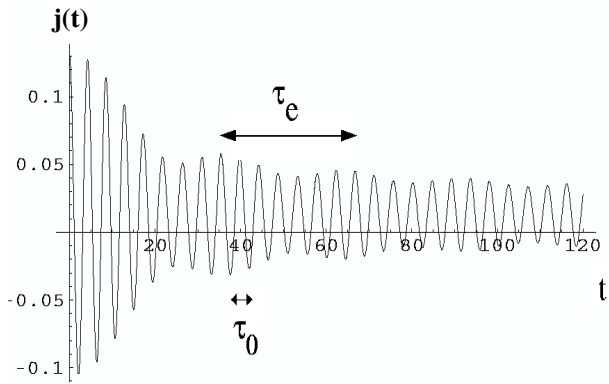}
\caption{Current evolution for an abrupt quench, $T=0$. We set
super-lattice strength $V_0/w=0.7$ and supercurrent boost
  $\nu={\pi\over 15}$. \label{Abrupt Approx fig}}
\end{figure}

Fig. \ref{Abrupt Approx fig} shows an example of the current evolution
for an abrupt quench. Generically, the evolution is characterized by fast oscillations
superimposed on a slow envelope which oscillates and decays.
The time between successive maxima of the slow envelope can be obtained
by looking at the extrema of the Fresnel integrals in Eq. \eqref{Def
M}. From $\partial_z {\cal C}(z)=\cos(z^2\pi/2)$, we see that the
extrema are at $z=\sqrt{2n}$, with $n$ integer. Thus the condition
$\sqrt{2n}=\sqrt{t}\nu\kappa$ yields the envelope period $\tau_e$:
\begin{eqnarray}
\label{envelope period}
\tau_e={4\over \nu^2\kappa^{2}}.
\end{eqnarray}
The fast current oscillations are a consequence of Bloch-like oscillations
in each of the 2-state systems (k-state fermions and their $k-\pi$
backscattered partners) described by Eq. \eqref{H k}. These arise
from the $\cos(2t-T)V_0$ terms in Eq. \eqref{current envelope
Rosen Zener}, and thus have period $\tau_B={\pi\over V_0}$. The
Bloch oscillations decohere over time due to the spread in
frequencies of the 2-level systems, given in Eq. \eqref{E k}.
Indeed, we note that the current in Eq. \eqref{current envelope
Rosen Zener} eventually decays, and no current is left in the
system, to lowest order in $\nu$.

Nevertheless, as the system is integrable and not coupled to a
heat bath, some current survives at higher order in $\nu$. By
computing the current averaged over long times,
${\overline{\langle j\rangle}}=\lim_{{\cal
T}\rightarrow\infty}{1\over {\cal T}}\int_0^{\cal T}{\langle
j\rangle}\D t$,  we show that the surviving current is
proportional to $\nu^3$ for small supercurrents. From Eq.
\eqref{current integral}, we find that the averaged current is
given exactly by:
\begin{eqnarray}\label{surviving current superlat}&
{\overline{\langle j\rangle}}={2w\sin(\nu) \cos(V_0T)\over \pi
}\Big(1-{V_0\arctan({2w\sin(\nu) \over V_0})\over
  2w\sin(\nu)}\Big),
\end{eqnarray}
which, to leading order in the supercurrent, is:
\begin{eqnarray}\label{surviving current small}
 {8w^2\nu^3 \over 3\pi
  V_0^2}\cos(V_0T),
\end{eqnarray}
which is of order $\nu^3$, and thus is absent from the treatment
leading to Eq. \eqref{current envelope Rosen Zener}.

%

To complement the current evolution analysis, we next analyze the
momentum distribution of HCBs undergoing the SF-insulator quench
while carrying a supercurrent. Although the analysis of the
current evolution in the HCB system is the same for free fermions
and a Tonks-Girardeu gas, the momentum distribution of the two
systems is quite different. While the presence of a supercurrent
is described by a shift of the Fermi step-function of the free
Jordan-Wigner fermions, as in Eq. \eqref{Fermi step supercurrent},
in terms of the bosons, the supercurrent leads to a peak in the
boson momentum occupation $n_k=\sum_l e^{ikl}\langle
b^{\dag}_l(t)b_0(t)\rangle$ at the boost value, i.e. $n_k\propto
|k-\nu|^{-1/2}$ \cite{Lenard}. Upon quenching the system away from
the critical superfluid, momentum modes at $k$ undergo Bloch
oscillations with modes at $k+\pi$, resulting in oscillations of
the current which eventually decays to an average, non-zero value.
Unfortunately, an analytic description of the momentum
distribution evolution is a much harder task, requiring detailed
analysis of the determinants arising in the boson correlation
functions. Even at equilibrium the analysis is not trivial, and
relies on the translational invariance of the system which allows
the application of mathematical machinery such as Szego limit
theorems. Our case is much less accessible, due to the incoherent
mixing between the different $k$ modes.

\begin{figure}
\includegraphics[scale=0.4,bb=105 214 610 478,clip]{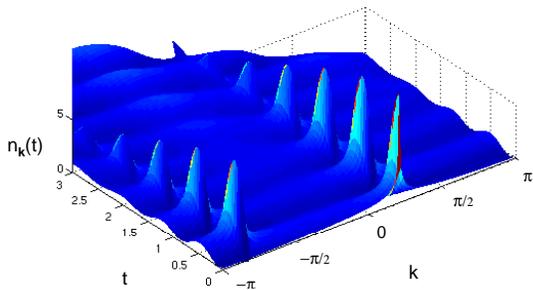}
\caption{$n_k$ as a function of $k$ and time for a boost of
  $\nu=\pi/5$, and super-lattice $V_0=4w$. The initial peak is at
  $k=\nu$, and the peak transports back and forth to $-4\pi/5$
  before disappearing due to decoherence. Results obtained using an exact
  diagonalization study of chains 150
  sites long.  \label{n-of-k}}
\end{figure}

\begin{figure}
\includegraphics[width=8.5cm,height=4cm,bb=170 296 686 510,clip]{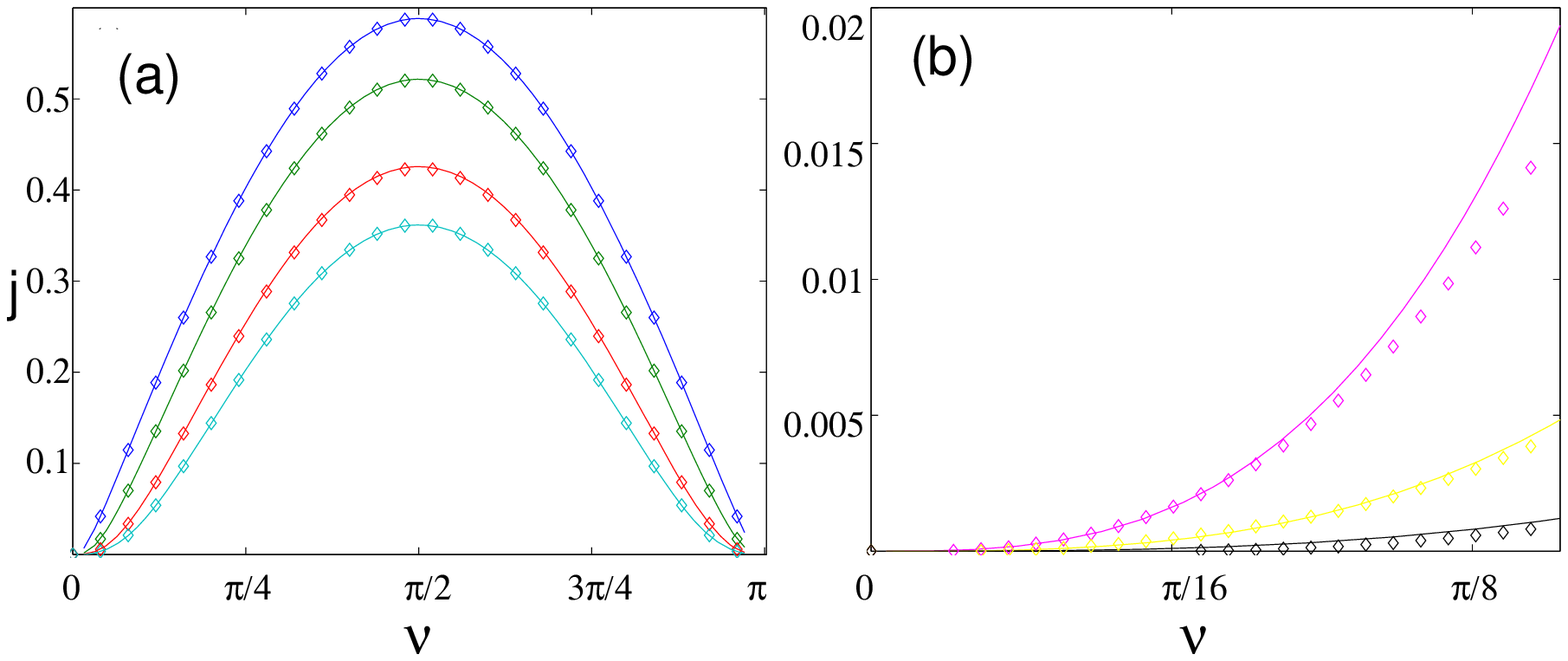}
\caption{Supercurrent surviving through a sudden quench. The symbols are the result of
exact diagonalization of a chain 350
  sites long.
(a) Current survival over the full range of boosts. From top to
bottom, $V_0/w = 0.1, 0.25 , 0.5, 0.7$ and the solid lines are
Eq.\eqref{surviving current superlat}. (b) The $\nu^3$ behavior of
current survival at small currents. From top to bottom,
 $V_0/w = 2, 4, 8$. The solid lines are Eq.\eqref{surviving current small} } \label{Isurvive}
\end{figure}

Using exact-diagonalization
techniques (see e.g. \cite{Rigol Muramatsu}) to monitor the time
evolution of the boson momentum distribution and the current, we
investigated lattice sizes of up to 350 sites. The momentum
distribution indeed reflects the supercurrent Bloch oscillations, as can be seen
 in the bosonic $n_k$ plotted in Fig. \ref{n-of-k}. The period of these oscillations of the peak in $n_{k}$ between $\nu$ and $\nu-\pi$
agrees very well with the analytical result $\pi\over V_0$. We also confirmed numerically the main results of
this manuscript, i.e., the current survival after a quench, Eqs. (\ref{surviving current superlat}) and (\ref{surviving
current small}). Our results for the long-time averaged
current (in a chain of 350 sites) vs. supercurrent boost $\nu$ and
super-lattice strength $V_0/w$ for an abrupt quench are summarized in
Fig. \ref{Isurvive}. Perfect agreement is obtained between the
analytical and numerical results.

In this paper we studied the behavior of hard-core bosons under a
Rosen-Zener quench in the presence of a supercurrent. We described
the evolution of the supercurrent, and its long-time survival
fraction, as well as the corresponding momentum distribution
evolution. Perhaps the most readily accessible result is the
`$\nu^3$ law': starting with supercurrent $\nu$, the surviving
current in the insulating phase is $\propto\nu^3$ for $\nu \ll 1$.
By using the Wigner-Jordan transformation, we essentially mapped
the HCB gas to a Fermi-system with back-scattering at the
Luther-Emery line \cite{Luther-Emery}. In light of our results, it
is particularly interesting to ask what happens when we consider a
Luttinger liquid (describing either fermions or bosons): is the
$\nu^3$-law universal, or does it depend on the Luttinger
parameter, $g\le 2$? This question, as well as a test of our
predictions and the level of their universality can be taken on
experimentally. This could be done, for instance, by probing
cold-atoms in a ring-shaped trap, with a superimposed optical
lattice. Such a geometry was recently discussed in Ref.
\cite{Rotating rings}. If the additional optical lattice is made
to rotate as it turns on, a supercurrent will exist in the
rotating frame. The insulating phase can then be accessed by
introducing a corotating super-lattice. Such an experimental setup
will also be able to probe nonequilibrium quenches well beyond the
regimes which we considered here analytically.

{\it Acknowledgment:} We would like to thank R. Santachiara and S.
Powell for discussions.

\end{document}